\documentclass{article}

\usepackage{arxiv}

\usepackage[utf8]{inputenc} 
\usepackage[T1]{fontenc}    

\usepackage{url}            
\usepackage{booktabs}       
\usepackage{amsfonts}       
\usepackage{amsmath}
\usepackage{nicefrac}       
\usepackage{microtype}      
\usepackage[title]{appendix}
\usepackage{graphicx}
\usepackage[bf]{caption}
\usepackage{float}
\usepackage{microtype}
\usepackage{subcaption}

\usepackage{doi}
\usepackage{etoolbox}
\usepackage[natbibapa]{apacite}
\AtBeginDocument{}
\renewcommand{\APACrefnote}[1]{}

\newtoggle{bibdoi}
\newtoggle{biburl}
\makeatletter
\newsavebox{\bib@url}
\newsavebox{\bib@doi}

\undef{\APACrefURL}
\undef{\endAPACrefURL}
\undef{\APACrefDOI}
\undef{\endAPACrefDOI}

\newenvironment{APACrefURL}
  {\global\toggletrue{biburl}\lrbox\bib@url}
  {\endlrbox}

\newenvironment{APACrefDOI}
  {\global\toggletrue{bibdoi}\lrbox\bib@doi}
  {\endlrbox}

\newcommand{\printinfo}{
  \iftoggle{bibdoi}{\usebox{\bib@doi}}{\usebox{\bib@url}}
  \togglefalse{bibdoi}
}

\AtBeginEnvironment{thebibliography}{
\pretocmd{\PrintBackRefs}{%
  \iftoggle{bibdoi}
    {\iftoggle{biburl}{\unskip\unskip}{}\usebox{\bib@doi}}
    {\iftoggle{biburl}{Retrieved from \usebox{\bib@url}}}{}
  \togglefalse{bibdoi}\togglefalse{biburl}%
}{}{}}
\makeatother

\title{Untangling The Relationship Between Power Outage and Population Activity Recovery in Disasters}

\date{} 					

\renewcommand{\headeright}{}
\renewcommand{\undertitle}{}
\renewcommand{\shorttitle}{}

\hypersetup{
pdftitle={Quantitative Measures for Integrating Resilience into Transportation Planning Practice: Study in Texas},
pdfsubject={},
}

\begin{document}
\maketitle

\begin{center}
{\Large
Chia-Wei Hsu\textsuperscript{a,*}, 
Ali Mostafavi\textsuperscript{a}
\par}

\bigskip
\textsuperscript{a} Urban Resilience.AI Lab Zachry Department of Civil \& Environmental Engineering, Texas A\&M University\\3136 TAMU, College Station, TX 77843\\
\vspace{6pt}
\textsuperscript{*} correseponding author, email: chawei0207@tamu.edu
\\
\end{center}
\bigskip
\begin{abstract}
Despite recognition of the relationship between infrastructure resilience and community recovery, very limited empirical evidence exists regarding the extent to which the disruptions in and restoration of infrastructure services contribute to the speed of community recovery. To address this gap, this study investigates the relationship between community and infrastructure systems in the context of hurricane impacts, focusing on the recovery dynamics of population activity and power infrastructure restoration. Empirical observational data were utilized to analyze the extent of impact, recovery duration, and recovery types of both systems in the aftermath of Hurricane Ida. The study reveals three key findings. First, power outage duration positively correlates with outage extent until a certain impact threshold is reached. Beyond this threshold, restoration time remains relatively stable regardless of outage magnitude. This finding underscores the need to strengthen power infrastructure, particularly in extreme weather conditions, to minimize outage restoration time. Second, power was fully restored in 70\% of affected areas before population activity levels normalized. This finding suggests the role infrastructure functionality plays in post-disaster community recovery. Interestingly, quicker power restoration did not equate to rapid population activity recovery due to other possible factors such as transportation, housing damage, and business interruptions. Finally, if power outages last beyond two weeks, community activity resumes before complete power restoration, indicating adaptability in prolonged outage scenarios. This implies the capacity of communities to adapt to ongoing power outages and continue daily life activities. These findings offer valuable empirical insights into the interaction between human activities and infrastructure systems, such as power outages, during extreme weather events. They also enhance our empirical understanding of how infrastructure resilience influences community recovery. By identifying the critical thresholds for power outage functionality and duration that affect population activity recovery, this study furthers our understanding of how infrastructure performance intertwines with community functioning in extreme weather conditions. Hence, the findings can inform infrastructure operators, emergency managers, and public officials about the significance of resilient infrastructure in life activity recovery of communities when facing extreme weather hazards. 

\end{abstract}



\section{Introduction}
\label{sec:Introduction}
Infrastructure services resilience is a key aspect for reducing impacts and accelerating community recovery within a disaster response plan \citep{miles_role_2012}. Critical infrastructure systems are the bedrock of a functioning society. Resilience refers to the ability of infrastructure services to withstand and quickly recover from disruptions or failures. Among critical infrastructures, power infrastructure may be one of the most important as it can have a significant impact on different perspectives including demographic, socioeconomic, transportation and communities \citep{esmalian_empirical_2020,ulak_assessment_2018}. Power outages caused by natural disasters can disrupt essential services, such as healthcare \citep{adhikari_earthquakes_2017,hiete_scenariobased_2011}, transportation \citep{ghorbanzadeh_city_2022}, and communication \citep{pourebrahim_understanding_2019}. Vulnerable populations such as the elderly, children, and low-income households may be disproportionately affected by power outages \citep{coleman_energy_2022,esmalian_disruption_2021,lee_community-scale_2022}. Additionally, power outages can exacerbate the effects of natural disasters by hindering emergency response and recovery efforts.\\
Despite the recognition of the relationship between infrastructure resilience and community recovery, very limited empirical evidence exists regarding the extent to which the disruptions in and restoration of infrastructure services contribute to the speed of community recovery. To address this gap, in this study, we evaluate the extent of impact and speed of restoration of power infrastructure and its contribution to the speed of population activity recovery (as an important aspect of community recovery) through the use of data collected from the 2021 Hurricane Ida in the United States. Figure 1 shows the overview of the study workflow.

\begin{figure}[!ht]
	\centering
	\includegraphics[width=0.8\textwidth]{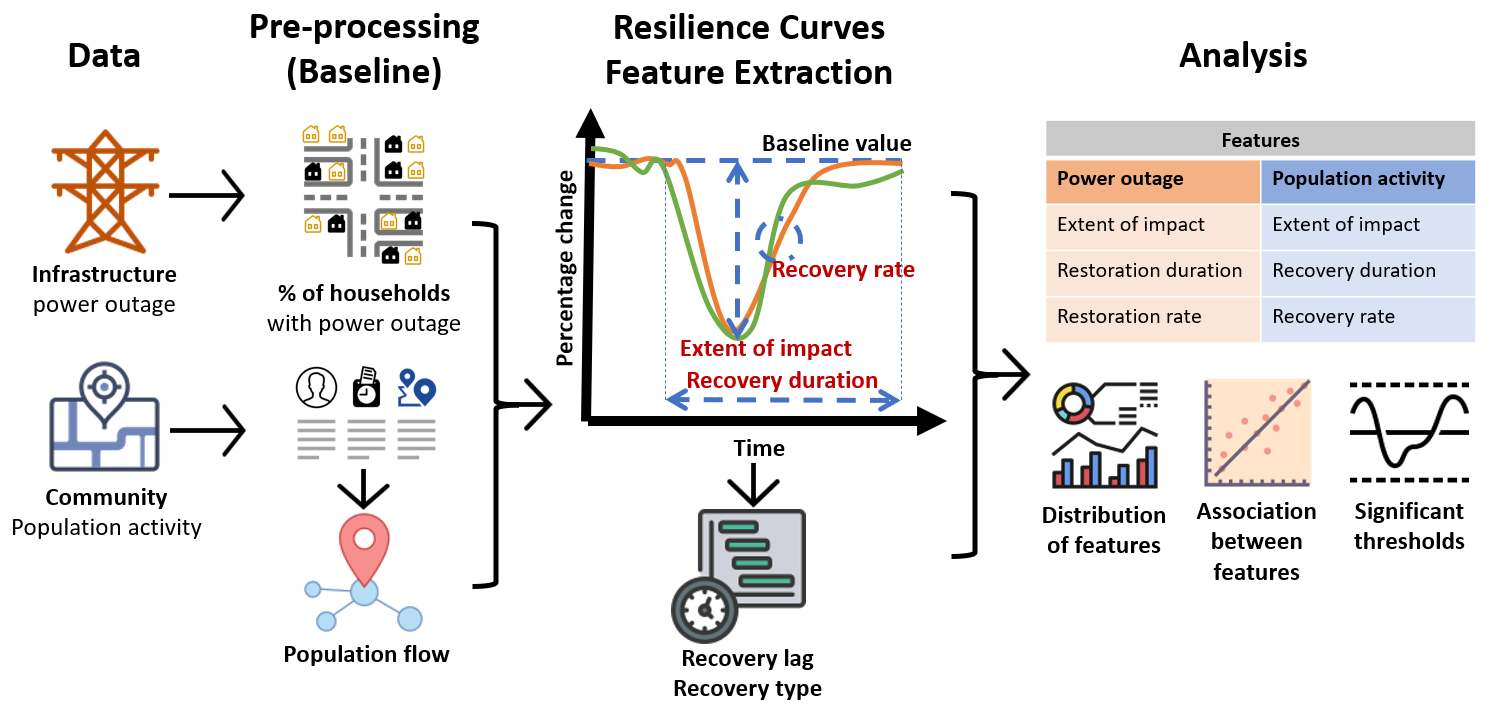}
	\caption{Overview of research framework: Location-based data and power-outage data are analyzed to estimate power outage and population activity as proxies for the performance of infrastructure system and communities under impact (of Hurricane Ida, in this study). Features, including the extent of impact, recovery/restoration duration, and recovery rate, are extracted from the resilience curves constructed for power outage and population activity; recovery lag and recovery type (whether infrastructure system and community restore/recovery first) can be obtained by comparing the recovery duration between the resilience curves. The distribution of the features collected, the association between features, and the possible threshold that governs the recovery types are examined in the analysis.}
	\label{fig:fig1}
\end{figure}

\section{Literature review}
\label{sec:Literature review}

In this section, we examine the existing literature in three related areas to highlight the point of departures for the current study.

\subsection*{Use of human mobility or population activity analysis during disasters}
Understanding human mobility and population activity during disasters can inform  effective disaster management and response. Recent studies have utilized social media data, mobile phone data, and other location-based data sources to analyze human mobility during disasters \citep{yabe_mobile_2022,hsu_human_2022-1}. These studies have shown that of population activity can be perturbed at different levels during hazard events \citep{jiang_data-driven_2023,wang_aggregated_2017}. Furthermore, human mobility data has revealed disparities in disaster evacuation patterns, with race and wealth playing a significant role \citep{deng_high-resolution_2021}. \cite{roy_quantifying_2019} quantified human mobility resilience to extreme events using geo-located social media data, emphasizing the importance of understanding mobility patterns during disasters for effective disaster response and recovery. \cite{hsu_human_2022-1} also found that human mobility networks manifest dissimilar resilience characteristics at different scales, highlighting the complexity of human mobility during disasters. The review of literature shows that the extent of impact and recovery of population activities during hazard events can be effectively quantified and analyzed as an important dimension of community recovery analysis \citep{lee_specifying_2022,yuan_smart_2022}. Hence, in the current study, we utilized human mobility data to evaluate the extent and speed population activity recovery.

\subsection*{Importance and impact of power outage on communities during hazard events}
Power outages during hazard events have significant consequences on communities, affecting a wide range of services and sectors, including healthcare, transportation, and communication \citep{munzberg_spatial-temporal_2017}. A study by \cite{coleman_energy_2023} highlights the presence of energy inequality in climate hazards based upon the social and spatial disparities in managed and hazard-induced power outages. This energy inequality can disproportionately affect vulnerable communities, exacerbating the difficulties faced during hazard events \citep{lee_community-scale_2022,moreno_community_2019}. Similarly, \cite{li_leveraging_2020} leveraged social media data to study community resilience in New York City during the 2019 power outage. These studies highlight how communities would be affected by power outages, regional and population group differences, and the need for preparedness to minimize negative impacts. Despite recognition of the importance of resilience power infrastructure, very little empirical evidence exists regarding the relationship between the influence of extent of power outages and speed of power restoration and community recovery. 

\subsection*{Relationship between infrastructure resilience and community recovery}
Infrastructure resilience plays a crucial role in community recovery following disasters \citep{miles_role_2012}. \cite{comes_measuring_2014} showcased the importance of resilient infrastructure for minimizing the negative effects of disasters on communities measured disaster resilience by investigating the impact of Hurricane Sandy on critical infrastructure systems. \cite{yu_quantifying_2019} demonstrated the link between infrastructure resilience and community recovery using hierarchical Bayesian kernel methods in a case study on recovery from power outages.  A comprehensive review by \cite{koliou_state_2020} on the state of research in community resilience highlighted the progress and challenges in understanding the relationship between infrastructure resilience and community recovery. The authors emphasized the need for further research to better understand how to enhance community resilience through improvements in infrastructure systems. \cite{yabe_resilience_2021} used large-scale mobility data to model recovery dynamics of interdependent urban socio-physical systems, providing valuable insights into the resilience of communities in the face of disasters. While these studies emphasize the importance of infrastructure resilience for community recovery and present analytical methods for analyzing the interplay between infrastructure systems and community functionality, limited empirical understanding exists regarding the relationship between the extent and duration of power outages and the trajectory of population activity recovery in extreme events.

\section{Study Context}
\label{sec:Study Context}
On August 29, 2021, Hurricane Ida made landfall near Port Fourchon, Louisiana, as a Category 4 storm with sustained winds of 150 mph. Hurricane Ida was one of the deadliest U.S. hurricanes, causing an estimated \$75 billion in damage, making it the fourth costliest Atlantic hurricane to make landfall in the United States. With storm surges and intense rainfall and wind damage, power outages persisted for weeks after the hurricane and 91 human fatalities were recorded in Louisiana \citep{hanchey_notes_2021}. The coastal region of Louisiana was particularly affected., More than 22,000 power poles, 26,000 wire spans, and 5,000 transformers being knocked over, broken, or destroyed. Hospitals struggled to provide services with backup generators and residents suffered from the summer heat. We collected data related to power outages and population activities in eight parishes in the New Orleans metropolitan area most affected by the hurricane: Jefferson, Orleans, Plaquemines, St. Bernard, St. Charles, St. James, St. John the Baptist, and St. Tammany. In Louisiana, parishes are the jurisdictional entity corresponding to counties in other states. Within this area, we collected power outage data and location based mobility data for further analyses.

\section{Data Description}
\label{sec:Data Description}
\subsection*{Power outage data}
Real-time power outage data from Hurricane Ida was gathered from PowerOutage.US and the Entergy power company website, with a focus on outages caused by Entergy, the primary energy provider for the affected areas in coastal Louisiana, serving 1,275,873 residents. Data collection included the total number of customers impacted and the last updated time, from August 29 through November 23, 2021. The frequency of data collection varied, with hourly recordings from 8:00 a.m. to 12:00 a.m. from August 29 to September 3; recordings every three hours from September 4 to October 25; and recordings at only 10:00 a.m., 4:00 p.m., and 9:00 p.m. from October 26 through November 23, due to the decrease in the number of noticeable outages.\\
The percentage of outages at each Zip code was calculated by dividing the number of impacted customers, obtained from the Entergy website, by the total population at each Zip code. It's worth noting that the Entergy website does not differentiate between commercial and residential customers. As the researchers could not obtain the total number of customers per Zip code, the outage data was normalized by the total population to prevent inaccurately inferring a higher number of outages in areas with larger populations.\\
To further validate the power data from the Entergy website, we compared it to the percentage of affected customers at the county level using data from PowerOutage.US. The county-level data from PowerOutage.US matched the Entergy data, with both showing similar patterns for changes in affected customers on the same day. We merged the Entergy data, which only reported Zip codes with at least one affected customer, with all Zip codes of relevant parishes. We considered only parishes where Entergy supplied at least 90\% of the total accounted customers according to PowerOutage.US.

\subsection*{Human mobility data}
This study utilized a GPS location dataset obtained by Spectus from smartphones. Spectus gathers large-scale anonymous location information from almost 70 million mobile devices in the United States through a compliant framework when users opt in to location services of partner apps. Users have the option to opt out at any time. Spectus works with more than 220 mobile apps that use its proprietary software development kits. Spectus collects data from about one in four smartphones in the United States, representing nearly 20\% of the population. The data is collected through partner apps and utilizes the internal GPS hardware of devices. GPS sensor log data has been used in the past for studying human mobility and travel mode detection due to its high spatiotemporal resolution. Spectus de-identifies data and applies additional privacy measures, such as obscuring home locations at the census-block group level and removing sensitive points of interest. The device-level data includes anonymized individual location information, ID, location, and time. Spectus provides access to its own data, tools, and location-based datasets from other providers through a data cleanroom environment for privacy-preserving analyses of multiple geospatial data assets.\\
The data collected by Spectus is transformed into daily trip counts between origins and destinations. The initial data included only anonymized device ID, coordinates, and visit times. The locations visited by each device are determined by identifying the census tract polygon it falls into. The device's movement or trajectory is then determined based on the visit times. Daily trip counts between census tracts are determined by aggregating on a census-tract level. To match the spatial resolution of the power outage data, daily trip counts are further aggregated into the Zip code-level based on the HUD-USPS ZIP Code Crosswalk data provided by HUD’s Office of Policy Development and Research (PD\&R). This file allocates census tracts to Zip codes. Using daily trip counts, we created a human mobility network for further analysis. In this study, we are interested in the fluctuation of total mobility flow, so we aggregated inflows and outflows. Sometimes inflow and outflow of an origin and destination have different meanings; however, we found that separating inflow and outflow within our period of interest did not yield additional information because they are proportional to the total flow and the have the same flow pattern. The baseline period was August 19 through August 28, 2021. This period is considered a normal period without any perturbations caused by the hurricane. The mobility flow within this period is viewed as baseline performance. Comparing the mobility flow during hurricane perturbed period with the baseline performance gives us the resilience curve.

\section{Methodology}
\label{sec:Methodology}
The aim of this study is to examine the association between power outage extent and recovery duration and the population activity recovery to address the following research questions: (1) To what extent does the magnitude of impact and speed of restoration in power infrastructure contribute to the extent and speed of population activity recovery? (2) What characteristics/features could explain variations in the relationship between the extent of impact and restoration speed of power infrastructure with population activity recovery?\\ 
In the analysis, we calculated metrics of the extent of impact, recovery (restoration) duration, and recovery (restoration) rate for both power infrastructure and population activity in each spatial area. Another metric examined is the recovery lag, defined as the difference between the recovery (restoration) duration of the two systems (i.e., the between full power restoration and full population activity recovery). The metrics examined in this study are summarized in Table 1. To extract the metrics, we obtain the resilience curves that show the relationship between a system's performance and the magnitude of the disturbance or stress it is subjected to from the raw data \citep{hillebrand_decomposing_2018,poulin_infrastructure_2021}. The extent of impact on population activity can be inferred from mobility flow on a certain day compared to the baseline value of mobility flow during the unperturbed period before Hurricane Ida. We consider the baseline value of mobility flow as full performance (100\%). For example, if the value of mobility flow observed on a certain day during Hurricane Ida is 80\% of the value of baseline value of mobility flow, then the performance loss or impact is 20\%. The impact on the power infrastructure, on the other hand, is determined by the percentage of households experiencing power outages. For example, when 70\% of households experienced power outages, then the performance loss or impact is 70\%. By applying the aforementioned definition and performing calculations, we transform the raw data into resilience curves for each area at the Zip code-level for both power infrastructure and population activity. From the resilience curves, we can find the turning points, as well as global maximums and minimums, and thus further calculate the extent of impact, recovery duration, and recovery rate. Combining these features with Zip code geodata provides insight into distributions and some spatial inferences.

\begin{table}[!ht]
    \caption{Metrics calculated for community and infrastructure systems extent of impact, recovery (restoration) duration, recovery (restoration) rate and recovery lag.\\}
    \centering
    \begin{tabular}{lll}
    \hline
     Systems & \textbf{Community} & \textbf{Infrastructure} \\
    \cline{2-3}
     & \textit{Population activity} & \textit{Power infrastructure} \\
    \hline
    Aspects & Extent of impact & Extent of impact \\
     & Recovery duration & Restoration duration \\
     & Recovery rate & Restoration rate \\
     & Recovery lag & Recovery lag \\
    \hline
    \end{tabular}
\end{table}

In the next step, we conduct decision-tree analysis to evaluate the relationship between the extent of power outages and power restoration speed with population activity recovery. The decision-tree analysis includes grid search to perform tuning with hyperparameters, including the maximum depth of the tree, minimum number of samples required to be at a leaf node’ and minimum number of samples required to split an internal node. We use GridSearchCV, a powerful tool in the scikit-learn library that allows searching for the best set of hyperparameters for a machine-learning model. It works by training the model multiple times with different combinations of the specified hyperparameters then evaluating the model's performance using cross-validation. The best set of hyperparameters is then chosen based on the highest performance score. GridSearchCV saves time and effort compared to manually tuning the hyperparameters, and it can improve the performance of our model \citep{myles_introduction_2004,song_decision_2015}. Based on the decision tree and thresholds that split the branches of the decision tree, we examine whether different extents of power outage and restoration speed would explain variations in population activity recovery duration of different spatial areas.

\section{Results}
\label{sec:Results}
\subsection*{Extent of impact, recovery duration and the recovery types}
First, we examine spatial patterns of the extent of impact in population activity and power infrastructure. The indicator here is the maximum performance loss compared to baseline performance. Figures 2A and 2C show the spatial distribution regarding the extent of impact on population activity and power infrastructure. Larger negative values and darker colors represent areas suffering from larger impact, which are mostly coastal areas. Figure 2B compares the difference between the extent of impact on both systems, population activity and power infrastructure. The mean performance losses for power infrastructure and population activity is 48\% and 83\%, respectively.

\begin{figure}[!ht]
    \centering
    \begin{subfigure}{0.32\textwidth}
        \centering
        \includegraphics[height=4.2cm,keepaspectratio,width=\textwidth]{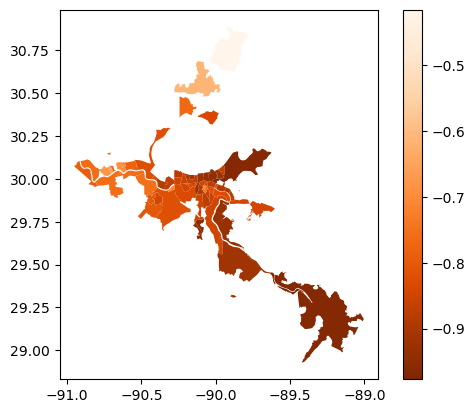}
        \caption{}
    \end{subfigure}
    \begin{subfigure}{0.32\textwidth}
        \centering
        \includegraphics[height=4.5cm,keepaspectratio,width=\textwidth]{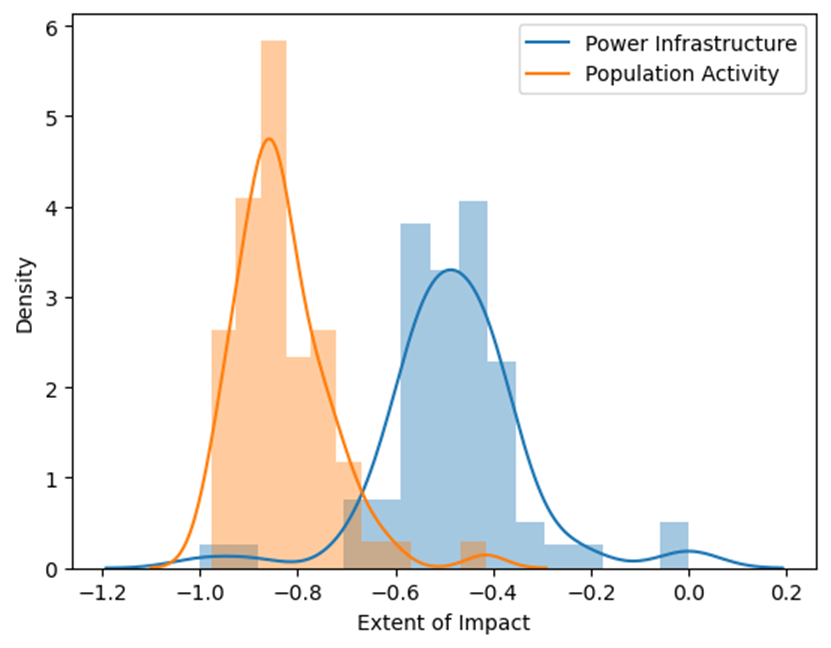}
        \caption{}
    \end{subfigure}
    \begin{subfigure}{0.32\textwidth}
        \centering
        \includegraphics[height=4.2cm,keepaspectratio,width=\textwidth]{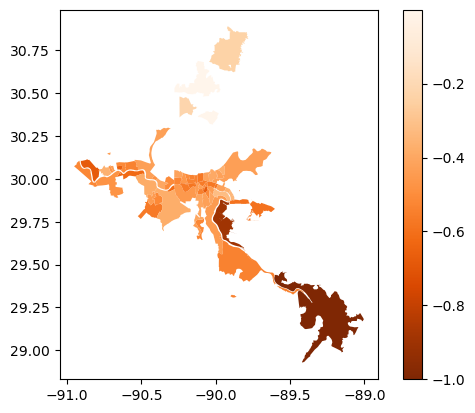}
        \caption{}
    \end{subfigure}
    
    \caption{(a) The spatial distribution regarding the extent of impact on population activity; (b) Comparison between the extent of impact on population activity (orange) and power infrastructure (blue); (c) The spatial distribution regarding the extent of impact on power infrastructure. The average extent of impact on population activity and power infrastructure are 48\% and 83\%, respectively.}
    \label{fig:fig2}
\end{figure}

Next, we examine the recovery (restoration) duration for both systems. Figures 3A and 3C show the spatial distribution of the recovery duration for population activity and power infrastructure. Larger values and darker colors represent areas experiencing longer recovery durations, which are mostly coastal areas. Figure 3B compares the distribution of the recovery (restoration) duration for both systems. The distribution for population activity is left skewed with a peak at around 20 days, while the distribution for power infrastructure recovery has a peak at 10 days. The average recovery duration of population activity and the restoration duration of power infrastructure are 17 days and 12 days, respectively. There is an observable offset of 5 to 10 days between the two distributions. Generally, population activity takes longer than the power infrastructure to fully recover. Anticipating power outages, though, residents might evacuate ahead of the hurricane landfall and return later after full power restoration.

\begin{figure}[!ht]
    \centering
    \begin{subfigure}{0.32\textwidth}
        \centering
        \includegraphics[height=4.2cm,keepaspectratio,width=\textwidth]{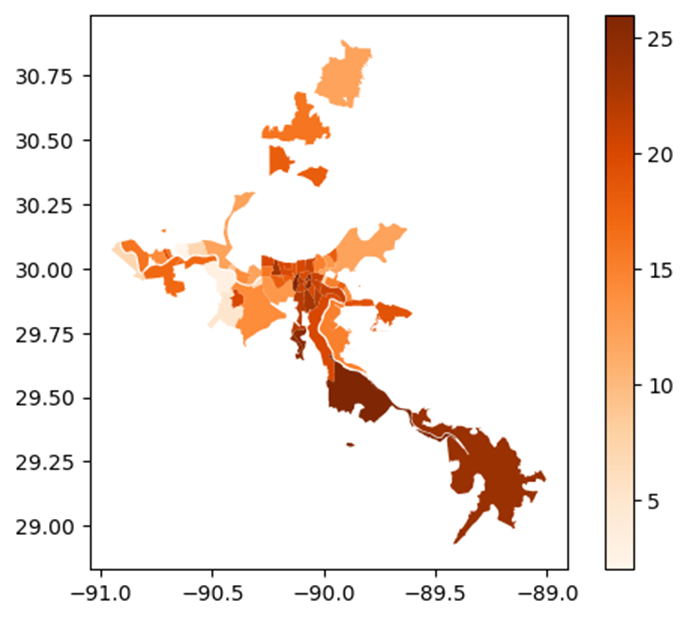}
        \caption{}
    \end{subfigure}
    \begin{subfigure}{0.32\textwidth}
        \centering
        \includegraphics[height=4.5cm,keepaspectratio,width=\textwidth]{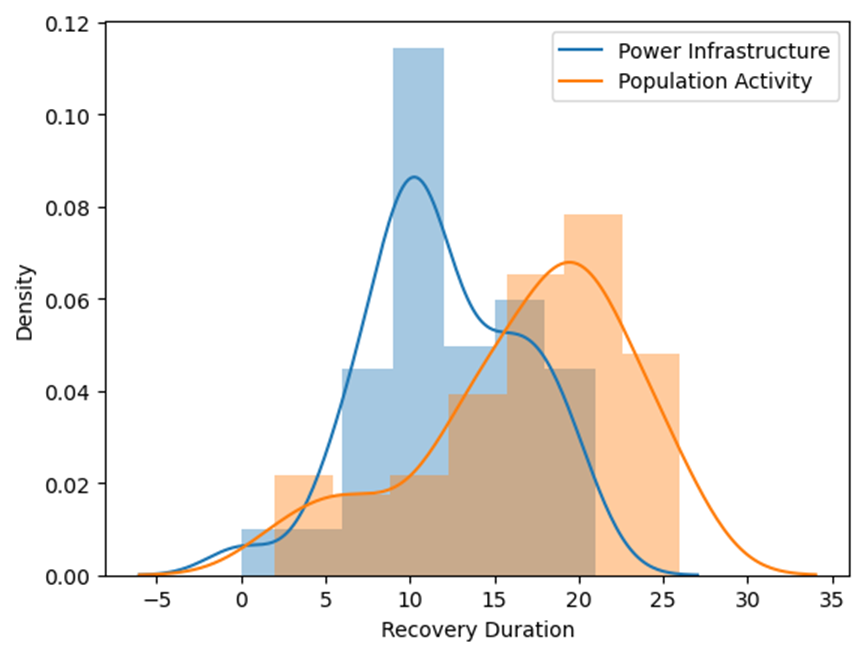}
        \caption{}
    \end{subfigure}
    \begin{subfigure}{0.32\textwidth}
        \centering
        \includegraphics[height=4.2cm,keepaspectratio,width=\textwidth]{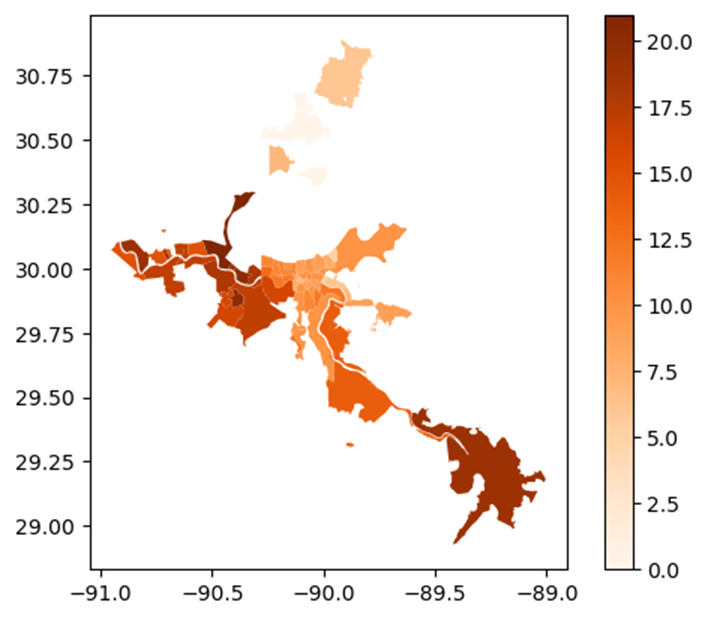}
        \caption{}
    \end{subfigure}
    
    \caption{(a) The spatial distribution regarding the recovery duration of population activity; (b) Comparison between the restoration duration of population activity (orange) and recovery duration of power infrastructure (blue); (c) The spatial distribution regarding the restoration duration of power infrastructure. The average recovery duration of population activity and the restoration duration of power infrastructure are 17 days and 12 days, respectively.}
    \label{fig:fig3}
\end{figure}

Comparing the recovery duration of the two systems, the results show the precedence relationship between the full recovery of the two systems as well as the recovery lag. For example, if the duration of full power restoration and full population activity recovery for an area is 15 days and 22 days, respectively, we know that full power restoration comes before full population activity recovery for this area and the recovery lag is 7 days (22 minus 15 days). Figure 4A shows the distribution regarding which system reaches full recovery first. Figure 4B shows which areas having which system reaches fully recovery first. Around 70\% of the areas reach full power restoration before full population activity recovery. These are the coastal and city center areas. 

\begin{figure}[!ht]
    \centering
    \begin{subfigure}{0.32\textwidth}
        \centering
        \includegraphics[height=4.2cm,keepaspectratio,width=\textwidth]{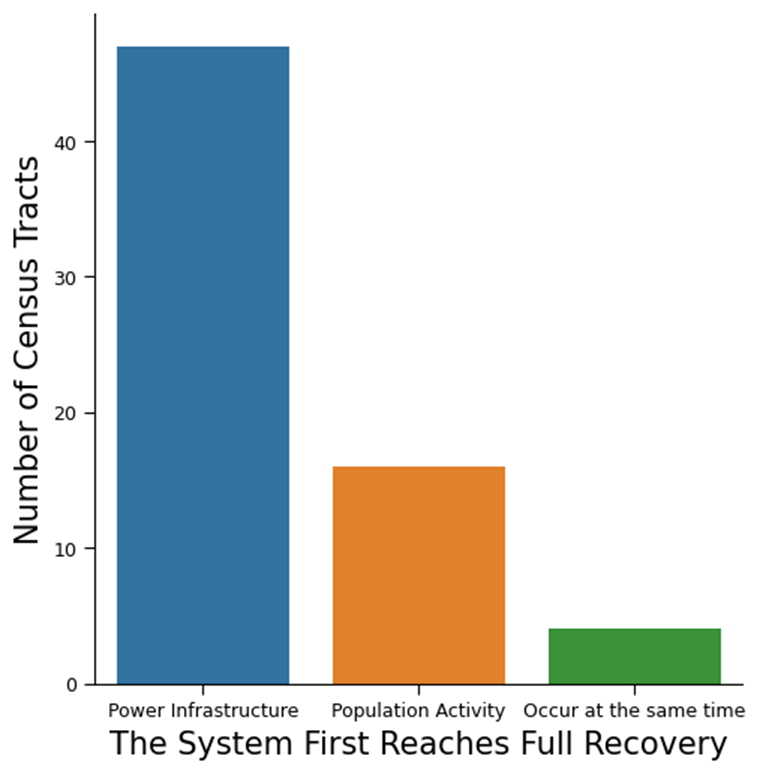}
        \caption{}
    \end{subfigure}
    \begin{subfigure}{0.32\textwidth}
        \centering
        \includegraphics[height=4.5cm,keepaspectratio,width=\textwidth]{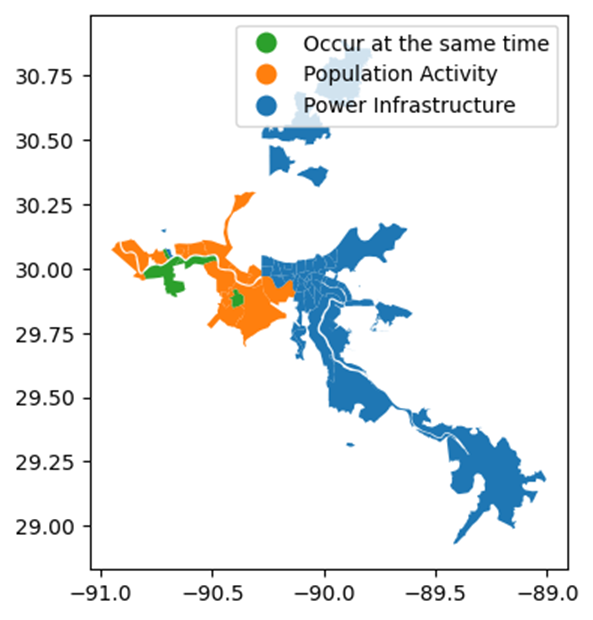}
        \caption{}
    \end{subfigure}
    
    \caption{The distribution regarding which of the systems (population activity and power infrastructure) first reaches full recovery.}
    \label{fig:fig4}
\end{figure}

\subsection*{Association between the recovery of infrastructure and human systems}
A decision (regression) tree is applied to predict the recovery rate of population activity based on features such as extent of impact, restoration duration and restoration rate of power infrastructure. Through gird search across the sets of hyperparameters, we obtain the best performing hyperparameters: maximum depth of the tree set to 2, minimum number of samples required to be at a leaf node = 2 and minimum number of samples required to split an internal node set to 2. Figure 5 shows the resulting decision tree. At the root node, the decision tree splits into two branches based on the feature power restoration duration. If the power restoration duration is shorter than 14.5 days, the tree follows the left branch. If the power restoration duration is longer than 14.5 days, the tree follows the right branch. The left branch next splits based on the feature power restoration rate. If the power restoration rate is smaller than 18.3\% per day, the tree follows the left branch of the split. If the power restoration rate is larger than 18.3\% per day, the tree follows the right branch. The pathway results show that if power restoration duration is less than 14.5 days and power restoration rate is less than 18.3\% per day, population activity recovery rate would be 4.49\% per day. If the power restoration duration is longer than 14.5 days and power restoration rate is greater than 18.3\% per day, population activity recovery rate is 6.88\%/day. The right branch next splits based on power restoration duration. If the power restoration duration is between 14.5 and 18.5 days, population activity recovery rate is 13.87\% day. 

\begin{figure}[!ht]
	\centering
	\includegraphics[width=0.7\textwidth]{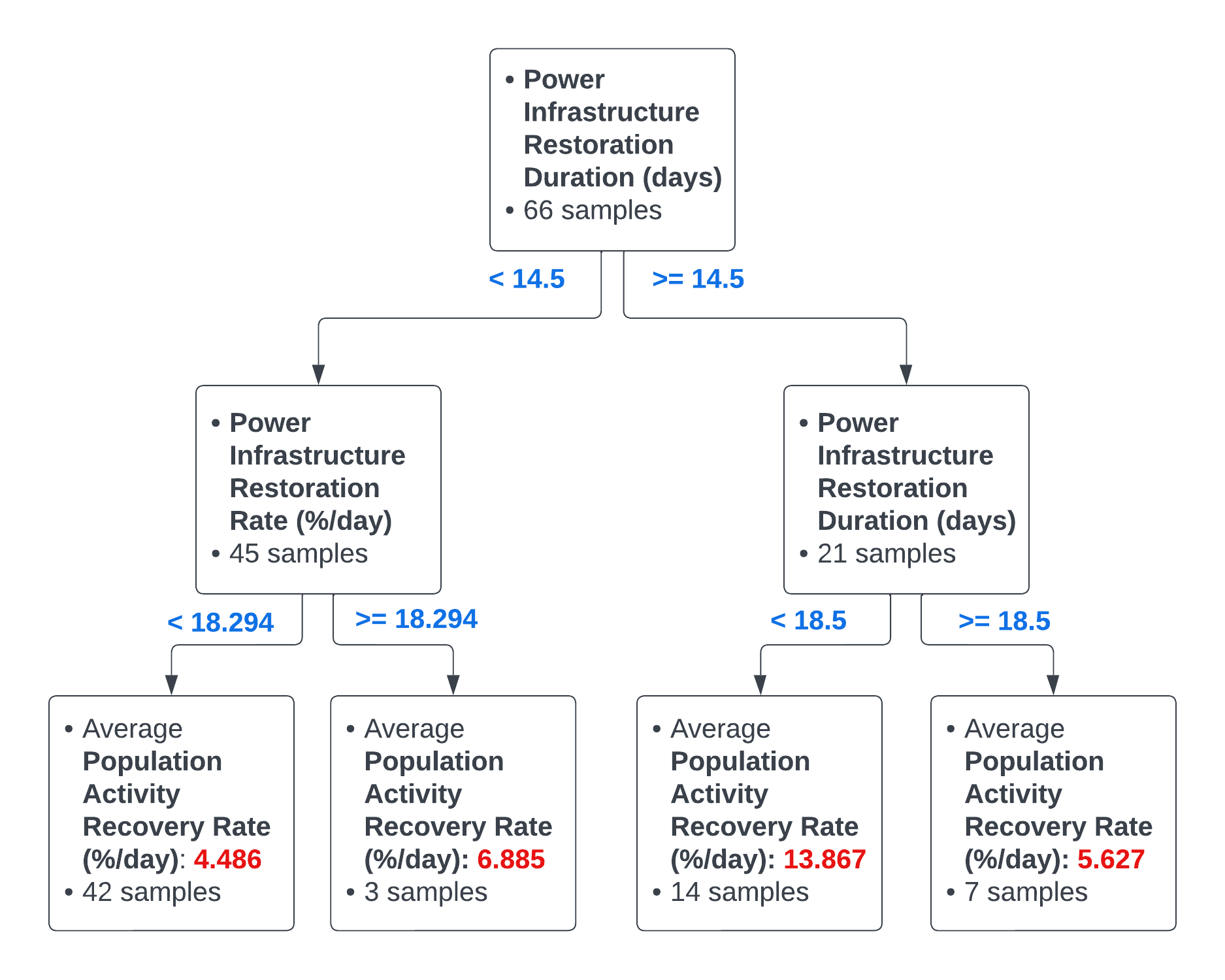}
	\caption{Decision (regression) tree predicting the recovery rate of population activity based on features such as the extent of impact, restoration duration and restoration rate of the power infrastructure. The bold black font represents the features used to split the branches of the decision tree. The bold blue font represents the thresholds that the sample in to groups. The bold red font represents the prediction on population activity recovery rate for each group.}
	\label{fig:fig5}
\end{figure}

The results show that, in areas with less than 14 days of power outage duration, in other words, the areas with faster rate of power outage restoration, the speed of population activity is higher. In areas with power outage durations greater than about 14 days, population activity recovery happens prior to full power restoration. We evaluate our decision tree by assessing the importance of each feature and the model’s performance in terms of its ability to accurately predict the target variable. The variable power infrastructure restoration duration alone accounts for 98.273\% of the importance for making predictions being the main feature of the model while power infrastructure restoration rate account for the rest of 1.727\%. Coefficient of determination (R-squared) measures the proportion of the variance in the target variable that is explained by the model, with a value of 1 indicating a perfect fit. The mean squared error (MSE), mean absolute error (MAE), and root mean squared error (RMSE) are measures of the average difference between the predicted and actual values of the target variable, with lower values indicating better performance. Power restoration duration is the most significant in explaining the variability in population activity recovery rate. The R-squared is 0.53, MSE is 7.40, MAE is 1.71, and RMSE is 2.72.\\
Figure 6 visualizes the decision tree (predicting the population activity recovery rate based on the extent of impact, restoration duration and restoration rate of the power infrastructure) on a two-dimensional plane. The solid line represents the first split, and the dotted lines represent the second split, creating subspaces that group the samples. Colors of the sample points indicates which system first reaches full recovery and their size represent our target value (population activity recovery rate). By interpreting the decision tree, we can conclude that there is an association between infrastructure resilience measured by the extent of impact, restoration duration, restoration rate of the power infrastructure and community recovery measured by population activity. In the first split, where restoration duration of power infrastructure is 14.5 days, the two major recovery types (population activity first reaches full recovery and power infrastructure first reaches full restoration) are separated. When power infrastructure is restored fast enough (duration within 14.5 days), a greater power restoration rate may further contribute to slightly faster population activity recovery rate. When power infrastructure restoration duration exceeds the threshold, the association between power infrastructure restoration rate and the population activity recovery rate becomes insignificant. The power restoration features and population activity recovery features are more significantly associated within threshold, otherwise untangled.

\begin{figure}[!ht]
	\centering
	\includegraphics[width=0.7\textwidth]{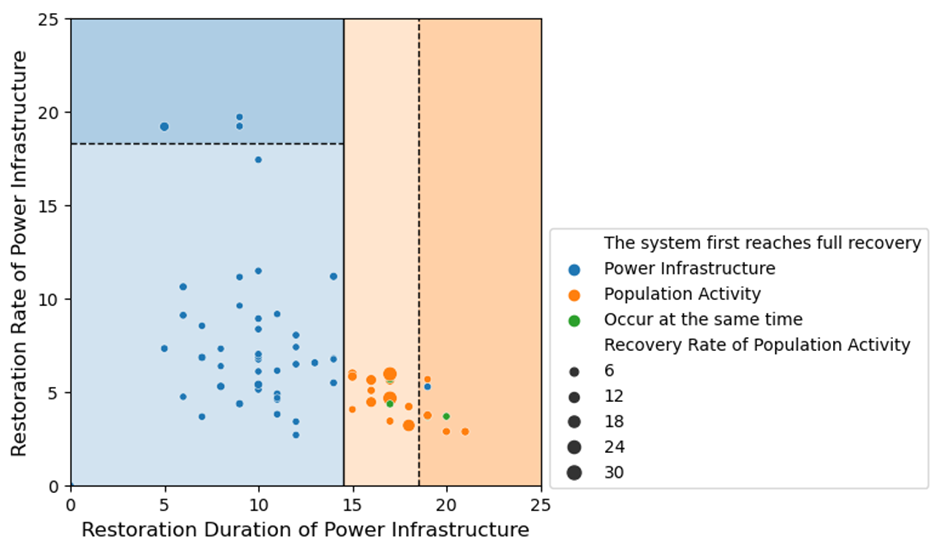}
	\caption{The two-dimensional visualization of the decision tree predicting the recovery rate of population activity based on extent of impact, restoration duration, and restoration rate of power infrastructure). The x-axis represents power infrastructure restoration duration and y-axis is the power infrastructure restoration rate. The solid line represents the first split (threshold: power infrastructure duration = 14.5 days) and the dotted lines represent the second split. The color of sample points indicates which of the systems first reaches full recovery and the size of sample points represents or target value (population activity recovery rate).}
	\label{fig:fig6}
\end{figure}

\subsection{Thresholds and notable resilience characteristics}
In the previous section, the decision tree provides several split rules or thresholds that separate data into groups that might share some characteristics regarding the recovery (restoration) speed. We also know that power infrastructure restoration duration and power infrastructure restoration rate are the more significant features for purposes of separating the data. Therefore, in the analysis presented in this section, we will discuss the interplay between these two features with the others.\\
There is a significant positive correlation between the extent of impact on power infrastructure and the restoration duration of power infrastructure (Figure 7). Generally, power infrastructure restoration duration increases when the extent of impact increases. The results also show that when the impact on power infrastructure exceeds a certain level, power restoration duration does not significantly increase. This level of impact can be characterized as a critical function loss level. If the extent of impact reaches this critical function loss level, the duration of restoration would not increase much beyond this point.\\
Figure 8 shows the relationship between the extent of impact on power infrastructure and the recovery lag between the two systems. There is a significant negative relationship between power infrastructure restoration duration and the lag between the full recovery of power infrastructure and population activity. We can clearly see that the recovery lag decreases as the power restoration duration increases. The recovery lag between the two systems is shorter when the restoration duration of power infrastructure is longer (>14.5 days). Population activity fully recovers slightly before the full restoration of power infrastructure. The recovery lag between the two systems is longer when the restoration duration of power infrastructure is shorter (<=14.5 days), yet sometimes power infrastructure fully restores much earlier than the full recovery of population activity. These findings reveal the significance of power system restoration in population activity recovery in the aftermath of extreme weather events.

\begin{figure}[!ht]
    \centering
    \begin{subfigure}{0.4\textwidth}
        \centering
        \includegraphics[height=4.5cm,keepaspectratio,width=\textwidth]{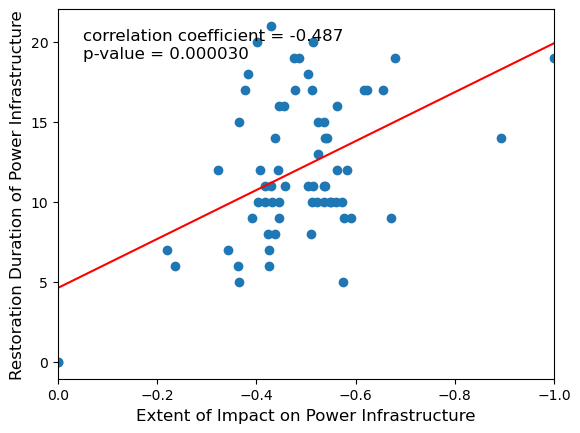}
        \caption{}
    \end{subfigure}
    \begin{subfigure}{0.4\textwidth}
        \centering
        \includegraphics[height=4.5cm,keepaspectratio,width=\textwidth]{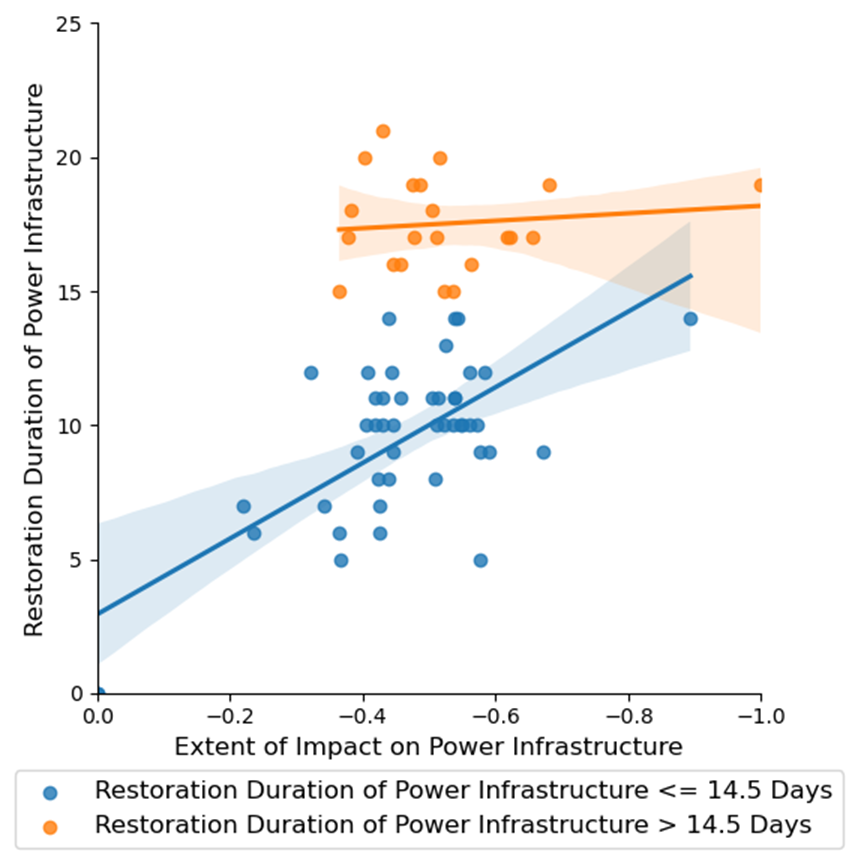}
        \caption{}
    \end{subfigure}
    
    \caption{Relationship between the extent of impact on power infrastructure and the restoration duration of power infrastructure. (a) The result shows that the two features are positively correlated: the greater the extent of outage, longer the restoration duration; (b) The result shows the data separated by threshold: power infrastructure restoration duration = 14.5 days. Full restoration of power infrastructure generally takes longer in areas suffering from greater impacts on power infrastructure.}
    \label{fig:fig7}
\end{figure}

\begin{figure}[!ht]
    \centering
    \begin{subfigure}{0.4\textwidth}
        \centering
        \includegraphics[height=4.5cm,keepaspectratio,width=\textwidth]{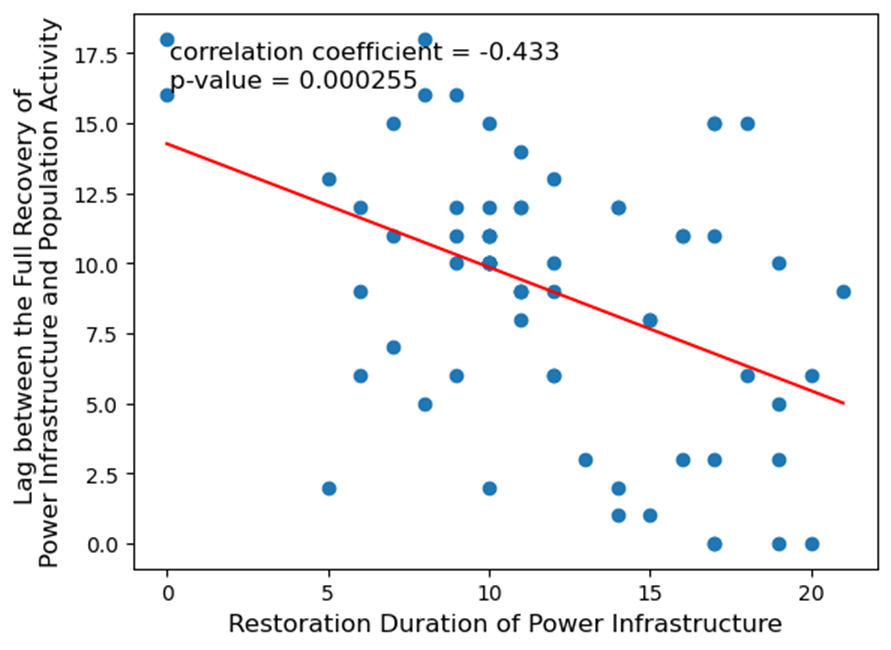}
        \caption{}
    \end{subfigure}
    \begin{subfigure}{0.56\textwidth}
        \centering
        \includegraphics[height=5cm,keepaspectratio,width=\textwidth]{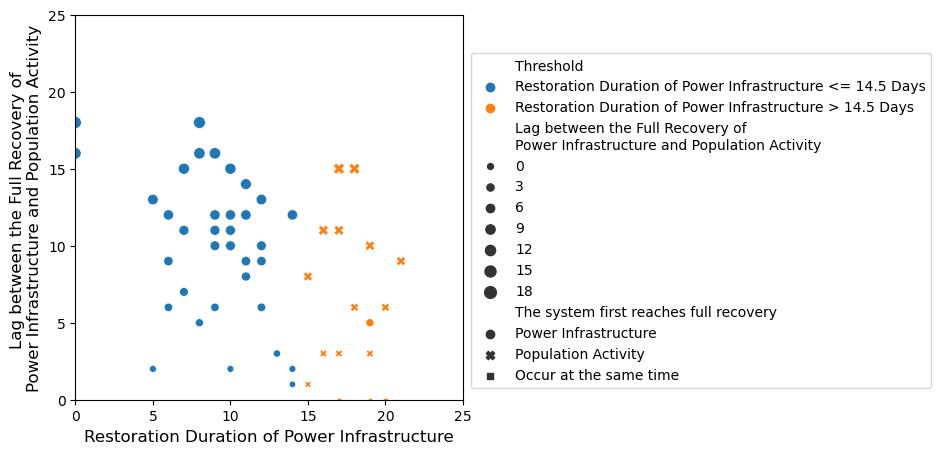}
        \caption{}
    \end{subfigure}
    
    \caption{Relationship between the extent of impact on power infrastructure and the restoration duration of power infrastructure. (a) The result shows that the two features are positively correlated: the greater the extent of outage, longer the restoration duration; (b) The result shows the data separated by threshold: power infrastructure restoration duration = 14.5 days. Full restoration of power infrastructure generally takes longer in areas suffering from greater impacts on power infrastructure.}
    \label{fig:fig8}
\end{figure}

\section{Discussion and Concluding Remarks}
\label{sec:Discussion and Concluding Remarks}
This study used empirical observational data to examine the relationship between infrastructure resilience and community recovery. Using metrics related to the extent of impact, recovery duration, recovery rate of population activity, and power infrastructure in the aftermath of Hurricane Ida, we evaluated the extent to which the extent of power outages and duration of power restoration shape the trajectory of population activity recovery.\\
The main findings of this study are threefold. First, the results show that the extent of power outages and the duration of power restoration are positively correlated if the extent of impact is less than a critical functionality loss. The critical functionality loss is the level of impact (i.e., extent of power outage) beyond which the duration of power restoration does not increase considerably with greater functionality loss. This result highlights the importance of strengthening the power infrastructure to reduce the extent of impact during extreme weather events in order to reduce the duration of power outage restoration. Second, the results show that for the majority (70\% of spatial areas) of impacted areas, full restoration of power was achieved before the population activity was fully recovered. This finding implies the important role of infrastructure functioning in the post-disaster recovery of communities. Also, the results show that rapid restoration of power is not a necessary condition for rapid population activity. In fact, the shorter the duration of power restoration, the longer the lag between population activity recovery and power outage restoration. This result is intuitive since population activity recovery also depends on other factors, such as transportation infrastructure impacts, housing damage, and interruptions to businesses and schools. Third, the results show that, if the duration of power outages exceeds more than 14 days, population activity recovery could occur prior to full power restoration. The 14-day duration can be a critical threshold beyond which power restoration and population activities untangle. This finding implies population adaptive capacity comes into play to cope with prolonged power outages, evidenced by people to returning to their day-to-day life activities even though power outages still persist.\\
These findings provide novel empirical evidence regarding the interplay between human systems (i.e., population activities) and infrastructure systems (i.e., power outages) in extreme weather events. Also, the findings provide a better empirical understanding of the relationship between infrastructure resilience and community recovery. Through characterization of critical functionality threshold for power outages and critical threshold for the duration of outages in explaining variations in population activity recovery, the findings of this study move us closer to better understanding of ways through which infrastructure performance interact with community functioning in extreme weather events. The findings can inform infrastructure owners and operators, emergency managers, and public officials regarding the ways through which resilient infrastructure could shape community recovery and the population’s life activities in disasters. Accordingly, in making resilience investments and infrastructure prioritization decision making, the loss of functionality of infrastructure can be better associated with disruptions and recovery of population activities to evaluate more objectively the benefits of such decisions.\\
In this study, the impact of hurricanes on population activity and power infrastructure were analyzed. The extent of impact, recovery duration, and recovery types found that the mean performance loss for power infrastructures and population activity was around 50\% and 90\%, respectively. Areas suffering from larger impact were mostly coastal areas. The recovery duration for both systems was also analyzed, with the power infrastructure showing a significant peak at 10 days and population activity showing a peak at around 20 days. There is an observable offset of 5 to 10 days between the two distributions. This results in population activity taking longer than power infrastructure to fully recover.\\
To investigate the association between the recovery of the systems, we applied a decision tree. A threshold of 14.5 days of restoration duration of power infrastructure was found to be the most significant. The variable restoration duration of power infrastructure alone accounts for 98.273\% of the importance for making predictions being the main feature of the model while restoration rate of power infrastructure account for the balance, 1.727\%. The decision tree provides good performance in terms of its ability to accurately predict the target variable, with power restoration duration being the most significant feature. When the extent of impact exceeds a certain threshold and restoration is not fast enough, population activity recovery starts regardless. In those areas, power and community recover at the same time or community may even recover before power restoration. When the power infrastructure takes more than 14 days to fully restore, people do not keep continue to wait, but rather resume life activities.\\
In conclusion, this study provides a comprehensive analysis of the impact of hurricanes on population activity and power infrastructure, and the association between the recovery of the systems. The results show that power restoration is an important factor contributing to the population activity recovery. Faster power restoration will help residents return more quickly to normal daily activity faster. Population activity is more affected by hurricanes than power infrastructure. The results of this study can be used to improve the resilience of power infrastructure and population activity in the face of hurricanes.

\section*{Acknowledgments}
\label{sec:Acknowledgments}
This material is based in part upon work supported by the National Science Foundation under CRISP 2.0 Type 2 No. 1832662 and the Texas A\&M University X-Grant 699. The authors also would like to acknowledge the data support from Spectus. Any opinions, findings, conclusions, or recommendations expressed in this material are those of the authors and do not necessarily reflect the views of the National Science Foundation, Texas A\&M University, or Spectus.

\section*{Author Contributions}
\label{sec:Author Contributions}
All authors critically revised the manuscript, gave final approval for publication, and agree to be held accountable for the work performed therein. C.H. was the lead Ph.D. student researcher and first author, who was responsible for supervising data collection, performed final analysis, and wrote the majority of the manuscript. A.M. was the faculty advisor for the project and provided critical feedback on the project development and manuscript.

\section*{Data availability}
\label{sec:Data availability}
All data were collected through a CCPA- and GDPR-compliant framework and utilized for research purposes. The data that support the findings of this study are available from Spectus, but restrictions apply to the availability of these data, which were used under license for the current study. The data can be accessed upon request submitted to the providers. The data was shared under a strict contract through Spectus’ academic collaborative program, in which they provide access to de-identified and privacy-enhanced mobility data for academic research. All researchers processed and analyzed the data under a non-disclosure agreement and were obligated not to share data further or to attempt to re-identify data.

\section*{Code availability}
\label{sec:Code availability}
The code that supports the findings of this study is available from the corresponding author upon request.

\bibliography{template}  






\end{document}